\documentclass[sigconf,screen]{acmart}

\usepackage[utf8]{inputenc} 
\usepackage[T1]{fontenc} 
\usepackage{microtype} 
\usepackage{makecell}
\usepackage{siunitx} 
\NewDocumentCommand{\anote}{}{\makebox[0pt][l]{$^*$}}
\usepackage{enumitem}   
\usepackage{comment}
\usepackage{multirow}
\usepackage{float}
\usepackage{tcolorbox}

\AtBeginDocument{%
  \providecommand\BibTeX{{%
    \normalfont B\kern-0.5em{\scshape i\kern-0.25em b}\kern-0.8em\TeX}}}

\setcopyright{acmcopyright}
\acmPrice{15.00}
\acmDOI{10.1145/3468264.3468595}
\acmYear{2021}
\copyrightyear{2021}
\acmSubmissionID{fse21main-p544-p}
\acmISBN{978-1-4503-8562-6/21/08}
\acmConference[ESEC/FSE '21]{Proceedings of the 29th ACM Joint European Software Engineering Conference and Symposium on the Foundations of Software Engineering}{August 23--28, 2021}{Athens, Greece}
\acmBooktitle{Proceedings of the 29th ACM Joint European Software Engineering Conference and Symposium on the Foundations of Software Engineering (ESEC/FSE '21), August 23--28, 2021, Athens, Greece}

\begin{document}

\title{A Grounded Theory of the Role of Coordination in Software Security Patch Management}


\author{Nesara Dissanayake}
\affiliation{%
  \institution{University of Adelaide}
  \city{Adelaide}
  \country{Australia}}
\email{nesara.madugodasdissanayakege@adelaide.edu.au}

\author{Mansooreh Zahedi}
\affiliation{%
  \institution{University of Adelaide}
  \city{Adelaide}
  \country{Australia}}
\email{mansooreh.zahedi@adelaide.edu.au}

\author{Asangi Jayatilaka}
\affiliation{%
  \institution{University of Adelaide}
  \city{Adelaide}
  \country{Australia}}
\email{asangi.jayatilaka@adelaide.edu.au}

\author{Muhammad Ali Babar}
\affiliation{%
  \institution{University of Adelaide}
  \city{Adelaide}
  \country{Australia}}
\email{ali.babar@adelaide.edu.au}


\begin{abstract}
  Several disastrous security attacks can be attributed to delays in patching software vulnerabilities. While researchers and practitioners have paid significant attention to automate vulnerabilities identification and patch development activities of software security patch management, there has been relatively little effort dedicated to gain an in-depth understanding of the socio-technical aspects, e.g., coordination of interdependent activities of the patching process and patching decisions, that may cause delays in applying security patches. We report on a Grounded Theory study of the role of coordination in security patch management. The reported theory consists of four inter-related dimensions, i.e., causes, breakdowns, constraints, and mechanisms. The theory explains the causes that define the need for coordination among interdependent software/hardware components and multiple stakeholders’ decisions, the constraints that can negatively impact coordination, the breakdowns in coordination, and the potential corrective measures. This study provides potentially useful insights for researchers and practitioners who can carefully consider the needs of and devise suitable solutions for supporting the coordination of interdependencies involved in security patch management. 
\end{abstract}



\maketitle

\section{Introduction}

Timely development and application of security patches to the identified vulnerabilities are considered critically important to avoid potentially successful security attacks \cite{souppaya2013guide}. Delays in patching security vulnerabilities can cause significant data losses, for example, the Equifax case \cite{Equifaxbug, Equifaxforbes}, or even human death \cite{Germancyberattack, Accenturereport2020}. Increased awareness about the potentially catastrophic consequences of delaying patching is leading to increased efforts aimed at improving technical and socio-technical aspects of software security patch management, hereafter called security patch management, a process that consists of identifying, acquiring, installing, and verifying security patches \cite{souppaya2013guide}. These activities entail several socio-technical aspects that underpin some of the critical decision points that make security patch management a complex and challenging undertaking \cite{li2019keepers, tiefenau2020security, huang2012patch}. Further, it needs the coordination of the efforts and decisions of multiple stakeholders with conflicting interests and several interdependencies. It has been reported that a majority of the delays in the security patch management process emerge from socio-technical aspects such as coordination \cite{li2019keepers, tiefenau2020security, nicastro2003security, dissanayake2020software}.

While it is widely understood that appropriate coordination support is needed for timely decisions and actions by the involved stakeholders, there is not much empirically known about the coordination aspects that may cause delays in applying patches. That means researchers and practitioners may not find much guidance in gaining a good understanding of the role of coordination in security patch management to answer some critically important questions: What is the role of coordination in security patch management? How does the coordination aspect cause delays in security patch management? What can be done for addressing the causes that may delay security patch management? These questions motivated us to carry out a grounded theory study of the role of coordination in security patch management.

In this paper, we present the first, to the best of our knowledge, Grounded Theory study exploring the role of coordination in security patch management. It is based on observations of 51 patch meetings over a period of 9 months, which involved 21 industry practitioners from two organisations in the mission-critical healthcare domain. We explain how coordination impacts the software security patch management process in four inter-related dimensions: causes, constraints, breakdowns, and mechanisms. Grounded in the evidence from industrial practices of the patch application decisions, the theory aspires to enhance the state-of-the-art understanding of researchers and practitioners in several ways: (a) the theory highlights the importance of gaining a deep understanding of the interdependencies before applying security patches and how improved support in coordination can help reduce the delays in security patch management; (b) it structures the knowledge about the unexplored phenomenon of security patching in the mission-critical healthcare domain; (c) provides a theoretical model to shape future Software Engineering (SE) research to address the practical concerns in security patching; (d) practitioners can leverage the understanding to reduce patching delays, and (e) the theory can also be useful to practitioners as guidance to enhance the confidence in patching decisions.


\section{Background and Motivation}
\label{section:background}

Software Security Patch Management is ``\textit{a multifaceted process of identifying, acquiring, installing, and verifying software security patches for products and systems}" \cite{souppaya2013guide, li2017large, tiefenau2020security}. Although there has been extensive research \cite{crameri2007staged, huang2012patch} on improving automation support in the security patch management process, we noticed a scarcity of empirical studies investigating the socio-technical aspects of security patch management. Existing empirical studies on socio-technical aspects of security patch management have primarily focused on studying system administrators \cite{crameri2007staged, dietrich2018investigating, li2019keepers, tiefenau2020security}, the patch management process and related challenges \cite{li2019keepers, tiefenau2020security}, and patch information retrieval behaviors and approaches of system administrators \cite{tiefenau2020security, jenkins2020anyone}.  

\hspace{1 mm} We found several studies (e.g., \cite{nappa2015attack, li2019keepers, huang2012patch, potter2005reducing, dissanayake2020software}) reporting coordination and collaboration challenges in the patch management process. However, they lack a comprehensive investigation of what causes the coordination needs and related challenges, its effects on the security patch management process, and the impact on delays of patch installation. Some studies \cite{nicastro2003security, prochazka2011race, hanauer2018process, tiefenau2020security} have described in high-level the dependencies between multiple stakeholders, such as vendors and organisations. Similarly, Nappa et al. \cite{nappa2015attack} reported that the coordination challenges concerning vendor dependencies arise from a lack of synchronized patch releases from different vendors because of shared vulnerabilities in the software code. Their analysis was based on a large data set of deployed patches in client-side vulnerabilities. Similarly, quantitative models and frameworks presented by a few other studies \cite{cavusoglu2008security, dey2015optimal, cavusoglu2006economics} focused on optimizing patch management by synchronizing the organisation's patch cycle with the vendor's patch release cycle to reduce patching costs and risks. As such, the reported dependencies with software vendors raise important concerns about the need for an in-depth understanding of the role and impact of such dependencies on security patch management.
    
\hspace{1 mm} However, coordination has been studied extensively across various dimensions in the related domains such as software development over the last decades \cite{cataldo2012coordination, strode2012coordination, bick2017coordination, cataldo2006identification, kraut1995coordination, van1976determinants}. The literature defines coordination as \emph{the management of interdependencies} \cite{crowston1994taxonomy, malone1991toward, malone1990coordination} and describes different types of coordination as explicit coordination and implicit coordination. Similarly, Crowston \cite{crowston1994taxonomy} provided a categorization of the types of dependencies based on the context such as task, knowledge, resource, and technical dependencies. Furthermore, the previous work \cite{bick2017coordination} about the coordination challenges in software development processes has demonstrated that ineffective coordination of dependencies represents a major cause of project failure, justifying the need for effective coordination to manage various interdependencies. Correspondingly, a comprehensive understanding of the role of coordination in patch application presents a critical research gap, which is fulfilled by this study.

\section{Research Method}
\label{section:method}

We used Grounded Theory (GT) \cite{glaser1967discovery, glaser1992basics} for data collection, analysis, theory development, and reporting. GT enables the systematic generation of theory from data, relating to social interactions and behaviour in real-world settings \cite{glaser1978theoretical, glaser1967discovery}. The choice of GT as our research method was based on two reasons:

\begin{enumerate}[label=(\alph*)]
    \item  The aim of our research, to understand the socio-technical aspects of security patch management in practice suited well with GT as it allows the investigation of people and interactions in a real-world phenomenon \cite{glaser1978theoretical}.
    \item  GT is considered most relevant to research areas that have not been deeply explored before \cite{hoda2011impact}, and research on the socio-technical aspects of security patch management is limited in the literature.
\end{enumerate}

We followed the Glaserian version of GT \cite{glaser1992basics} since it offers more flexibility to uncover the underlying concerns from the emergent data rather than limiting the research angle with a defined research hypothesis upfront. Following the guidelines, we started with an \emph{``area of interest"} - Socio-technical concerns in Security Patch Management. The guidelines by Stol et al. \cite{stol2016grounded} were followed for reporting the GT findings.  

\subsection{Data Collection}

We observed 51 patch meetings between two organisations (Alpha and Beta) in Australia, attended by 21 key stakeholders from 8 teams. These stakeholders represented diverse roles centered on decision-making, planning, and executing security patching. The longitudinal study was conducted from March 2020 - January 2021. The meetings were held every fortnight lasting approximately an hour and a half. Due to COVID-19, the meetings were held online through Microsoft Teams. The patch meetings focused on reporting security patch management status, tracking vulnerability remediation progress, discussing issues, planning patch cycles, and decision-making. Alpha is an Australian state government health services agency and Beta is an American multinational corporation that provides \textbf{security patch management} service to Alpha. While the most important and critical OS security patching was being outsourced to Beta, Alpha's in-house developed applications and other third-party applications were patched by Alpha teams and the respective third-party vendors. Table~\ref{tab:observationdetails} presents the investigated teams' demographics. Abiding by the human ethics guidelines, the details of the organisations and teams have been kept confidential.

We held discussions immediately after the meetings with one of Alpha's security team members to clarify any doubts that emerged during the observations and gather additional information. We also gathered data by analysing artefacts such as meeting minutes and patch mailing thread to supplement our understanding of the process, practices, and used terminology. 

The first author attended all 51 meetings that were held over the course of 9 months and conducted all 11 post-meeting discussions. All the meetings and discussions in Table \ref{tab:datacollectionsummary} were audio-recorded with permission and shared with all researchers. We adopted the protocol proposed by Spradley \cite{spradley2016participant} to guide the data collection during the meetings (see Appendix \ref{appendix:observationprotocol}).
The first author briefed other authors about the key aspects of the fortnightly meetings and the post-meeting discussions regularly. The data collection and analysis were performed in iterative and intertwined stages throughout. We continued with the data collection until the data analysis confirmed \emph{theoretical saturation}. The last few observations (M46-M51) provided more examples and evidence for the emerged findings during the analysis, but no new concepts, categories, or insights emerged. All authors mutually agreed that this was a clear indication of the \emph{theoretical saturation} and any additional data collection would not add value to the findings. 

\begin{table*}
  \caption{The investigated teams' demographics}
  \label{tab:observationdetails}
  \centering 
  \small
  \begin{tabular} {p{.08\textwidth}  p{.03\textwidth} p{.22\textwidth} p{.05\textwidth} p{.1\textwidth} p{.4\textwidth}}
    \toprule
    Organisation & Team & Domain & Team size \anote{$^*$} & Distribution \anote{$^+$} & Roles\\
    \midrule
    Alpha & T1 & Electronic Medical Records (EMR) & 5 & Co-located & EMR Application Owner, Server Engineer, System Administrator, Server Manager \\
     & T2 & Digital Health - Windows & 3 & Co-located & Server Engineer, System Administrator, Windows Application Specialist  \\
     & T3 & Digital Health - Non-Windows & 2 & Co-located & Unix Specialist, Server Engineer, System Administrator \\
     & T4 & Security & 1 & Co-located & Security Advisor \\
     & T5 & Change Management & 1 & Co-located & Change Manager \\
     & T6 & Clinical and Pathology Services & 1 & Co-located & Pathology Server Engineer \\
    \addlinespace
    \hline
    \addlinespace
     Beta & T1 & Server (Technical) & 7 & Distributed & Server Engineer, Senior Server Engineer, Unix Engineer, Server Manager, Client Delivery Manager \\
     & T2 & Finance and Audit (Non-technical) & 1 & Distributed & Accounts Manager\\
    \bottomrule
  \end{tabular}
\smallskip
\parbox[t]{\textwidth}{ $^*$~The team size refers to the number of team participants in the patch meeting.  $^+$~\textit{Distributed} refers to two locations within the state in Australia.}
\end{table*}

\begin{table}[htbp]
\caption{Summary of the data collection}
\label{tab:datacollectionsummary}
\centering 
\small
\begin{tabular}{ccccc}  
\toprule
\multirow{2}{*}{\parbox[c]{.15\linewidth}{\centering No. of meetings}}
 & \multirow{2}{*}{\parbox[c]{.1\linewidth}{\centering Duration}}
 & \multirow{2}{*}{\parbox[c]{.2\linewidth}{\centering No. of discussions}}
& \multicolumn{2}{c}{No. of hours} \\ 
\cmidrule{4-5}
 & & & {Meetings \anote{$^*$}} & {Discussions \anote{$^+$}}  \\
\midrule
51 & 9 months  & 11 & 30 hours & 7 hours \\
\bottomrule
\end{tabular}
\smallskip
\parbox[t]{0.5\textwidth}{ $^*$~The average time of a patch meeting=30 minutes.}
\parbox[t]{0.5\textwidth}{ $^+$~The average time of a post-patch meeting discussion=30-45 minutes.}
\end{table}

\subsection{Data Analysis}

We followed Glaser's data analysis procedure starting from \textbf{\emph{Open coding}} through \textbf{\emph{Selective coding}} to \textbf{\emph{Theoretical coding}} \cite{glaser1992basics, stol2016grounded, urquhart2012grounded}. 
The data analysis was led by the first author supported by other researchers who took the role of the validators at each stage throughout the iterative and intertwined rounds of data collection and analysis. All data including transcripts, observation and discussion notes, other artefacts (meeting minutes and patch mailing thread notes), codes, and memos were saved in the NVivo data analysis tool and shared with all co-authors. The second and third authors cross-validated all the emergent codes, concepts, categories, and core categories. Any conflicts in the coding and coding procedures were resolved in weekly detailed discussions between all authors throughout the analysis phase involving several rounds of revisions. Additionally, the emerged findings were further cross-checked with one of the senior members of Alpha’s security team.

\textbf{\emph{Open coding}} started with thoroughly reviewing the transcripts and recording \emph{key points} containing summarized phrases \cite{georgieva2008best}. It was further summarized into \emph{codes} of three-five words each, and any specific \emph{properties} of the code were captured in brackets, as shown in the example below.

\begin{tcolorbox}[colback=white, left=1pt, top=1pt, right=1pt, bottom=1pt]
    \textbf{Transcript:} \textit{``The patching is delayed because this .NET security vulnerability was reported in August after patching happened for the month. But, we also have another problem as this is a different version of .NET from what is standing across the fleet. This is .NET core, not .NET version 4.801."} \\
    \textbf{Key Point:} \textit{Need to identify and match Framework version dependencies} \\
    \textbf{Code:} \textit{Software application inter-dependencies (version)}
\end{tcolorbox}


Applying constant comparison on the codes that emerged within each observation, between different observations, and post-meeting discussions, we grouped them to a higher level of abstraction, i.e., \emph{concepts} \cite{glaser1967discovery, glaser1992basics}. Similarly, continuously comparing concepts produced \emph{categories}, a third-level of abstraction, and from categories generated \emph{core categories} \cite{glaser1992basics} at the end of the first round of coding. The core category represents the main problem or concern \textit{(core)} in the studied phenomenon, which presents the research question \cite{glaser1978theoretical, hoda2012developing}. Correspondingly, three potential \emph{core categories} emerged - Legacy software systems, Role of Coordination, and Role of Patch Meetings. The \% split of codes between the three core categories was 18\%, 51.3\%, and 30.7\% respectively. We selected \emph{``Role of Coordination in Software Security Patch Management"} as the core category because it met all the criteria defined by Glaser \cite{glaser1978theoretical} for selecting a category as the core. For example, the selected core category was central to other categories and frequently occurred in the data; meaningfully related to both other categories easily and took the longest to saturate. Our decision of focusing on the ``role of coordination" from the initial general focus on the ``role of the socio-technical aspects in security patch management" was informed by Glaser's Grounded Theory guidelines \cite{glaser1978theoretical}. 

\begin{figure*}[h]
  \centering
  \includegraphics[scale=0.85]{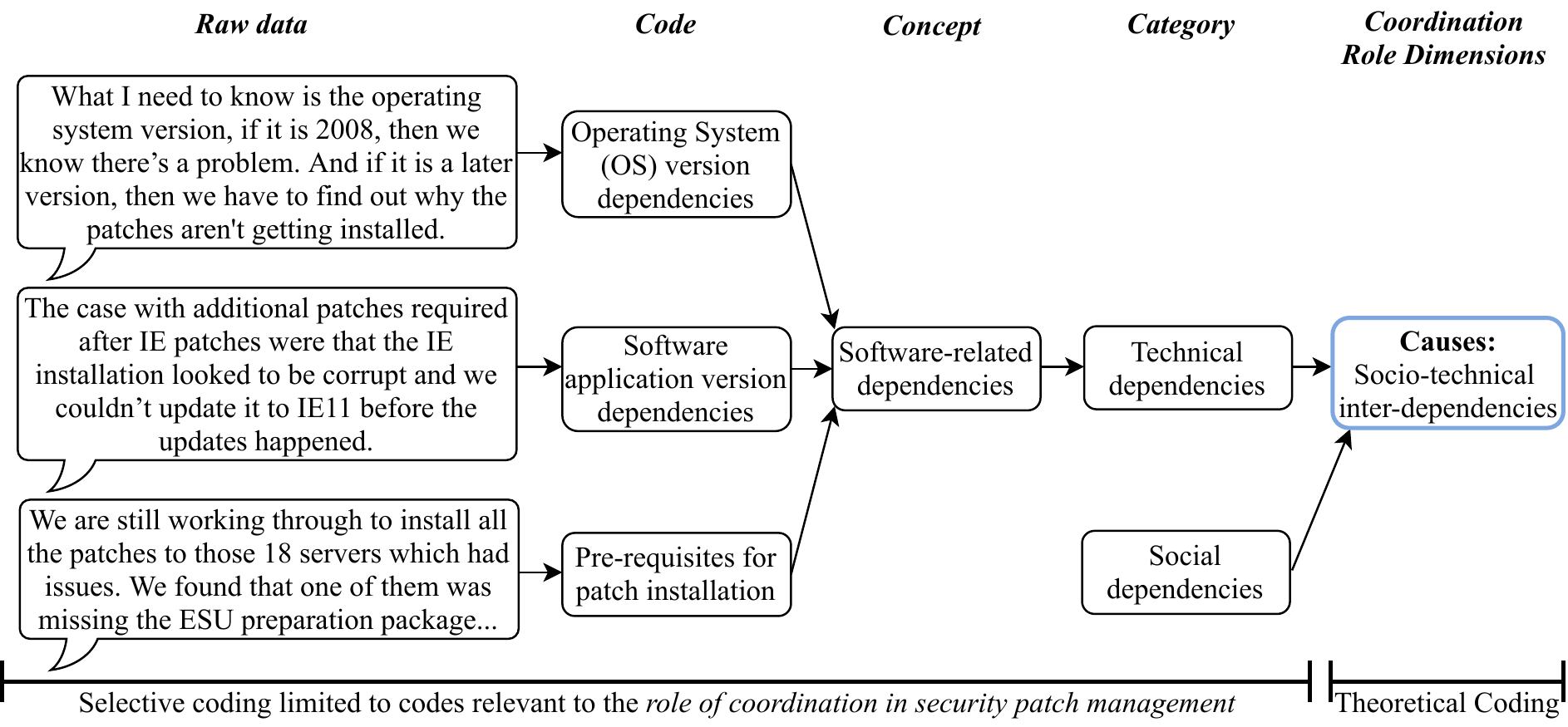}
  \caption{The emergence of category \textit{Socio-technical interdependencies} from the underlying codes and concepts.}
  \Description{Application of Glaser's data analysis procedure resulting in the development of the category \textit{Technical dependencies} during Selective Coding.}
  \label{fig:researchmethodcoding}
\end{figure*}

After establishing the core category, we continued \textbf{\emph{Selective coding}} \cite{glaser1978theoretical} limited to only those codes that were related to the selected core category. For example, Figure \ref{fig:researchmethodcoding} illustrates the emergence of the category \textit{Technical dependencies} that relate to the \emph{role of coordination} in the second round of coding - Selective coding. We continued to selectively code until no new insights or aspects emerged for each category, which indicated \emph{theoretical saturation} \cite{glaser1967discovery, glaser1992basics}.  

\begin{figure*}[h]
  \centering
  \includegraphics[width=0.875\textwidth]{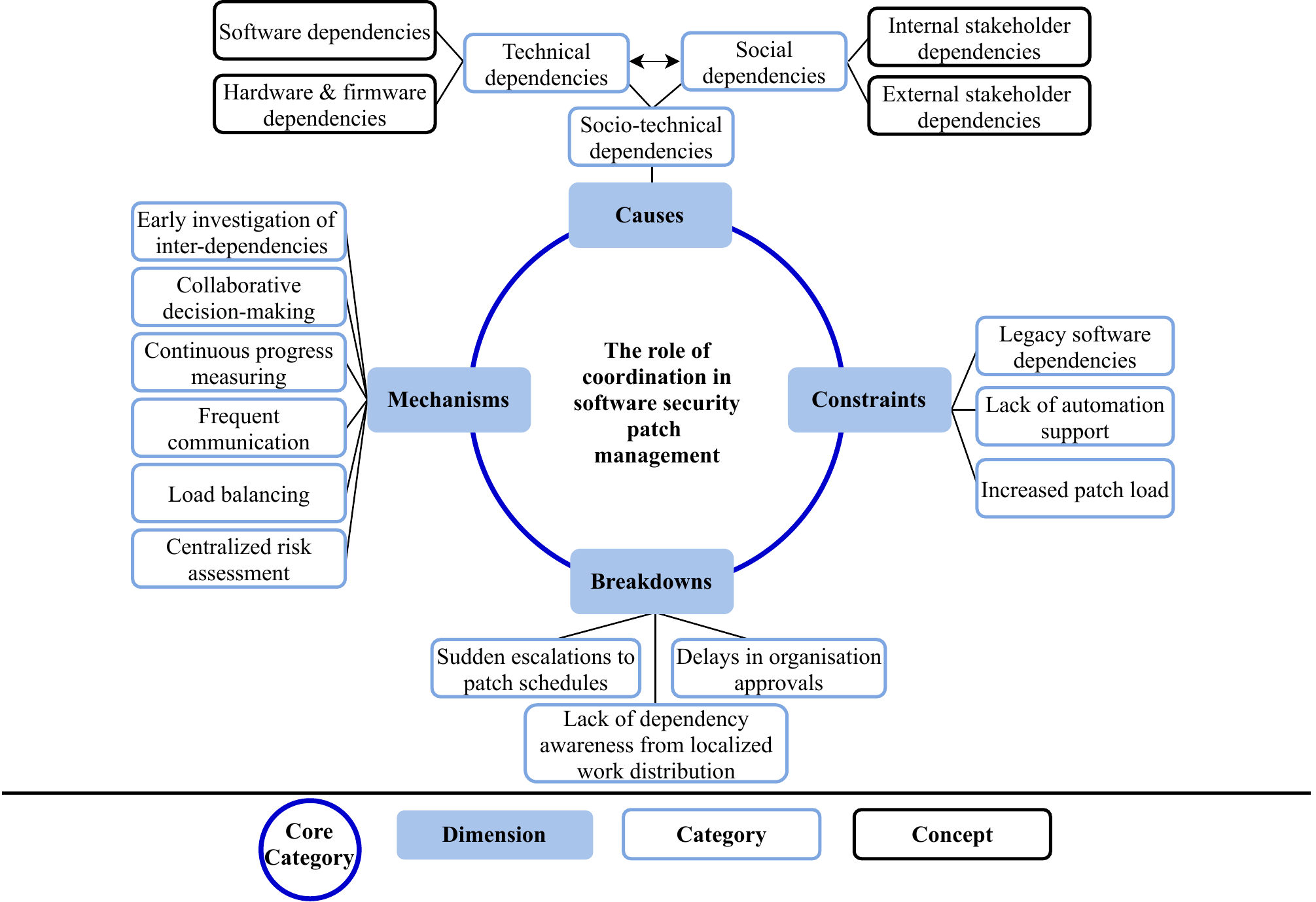}
  \caption{The theory of the role of coordination in software security patch management.}
  \Description{A grounded theory of the role of coordination in software security patch management described in four dimensions.}
  \label{fig:theory}
\end{figure*}

As the final step of the analysis, we applied \textbf{\emph{Theoretical coding}} \cite{glaser1978theoretical, glaser1992basics, glaser2005grounded, stol2016grounded} to establish conceptual relationships between categories, resulting in the development of a theory. We used \emph{memos} (\emph{memo sorting}) \cite{glaser1978theoretical, glaser1992basics} to guide us to uncover the links between the categories when developing the theory.
At this point, we consulted the literature, particularly Glaser's \emph{`theoretical coding families'} to find if any existing theoretical structure fits our current findings to visualize the theory. Glaser argues using a coding family to present the theory would help increase the completeness and relevance of the emerging theory \cite{glaser1978theoretical, glaser2005grounded}. As the findings (i.e., the causes, constraints, breakdowns, and mechanisms) emerged from the data, we found that the theoretical coding model, i.e., \emph{Dimension family} is the best fit to visualize the relationships between the categories. The \emph{Dimension family} is a theoretical structure that enables the findings to be presented as dimensions or elements of a phenomenon \cite{glaser1978theoretical, glaser2005grounded}, in this case, dimensions of the role of coordination in software security patch management. Thus, the theory of the role of coordination in security patch management, depicted in Figure \ref{fig:theory}, is described using: (a) \textbf{\emph{Causes:}} socio-technical dependencies that define the need for coordination; (b) \textbf{\emph{Constraints:}} factors that hinder coordination; (c) \textbf{\emph{Breakdowns:}} scenarios of patching failures resulting from ineffective coordination of the causes and constraints; and (d) \textbf{\emph{Mechanisms:}} strategies devised for supporting the coordination in security patch management.

\section{Findings}
\label{section:findings}

In this section, we describe the theory of the role of coordination in security patch management using four inter-related dimensions: \textit{Causes, Breakdowns, Mechanisms,} and \textit{Constraints}, providing evidence with grounded quotes from the underlying data. For ease of reference, we used unique identifiers to refer to participants, for example, \emph{AT4, M10-dis} refers to a participant from Alpha's Security team in the 10th post-meeting discussion, and \emph{BT1, M4} refers to a participant from Beta's Server team in the 4th meeting.

\subsection{Causes - Socio-Technical Dependencies}

We identified several socio-technical interdependencies inherent in the security patch management process that define the need for coordination. We recorded discussions about several scenarios exemplifying the potentially disastrous implications as a result of failing to identify dependencies before applying a patch. Through the data analysis process, we have placed the observed interdependencies in two categories: \emph{Technical} and \emph{Social} dependencies. 

\subsubsection{\textbf{Technical dependencies}}

The interdependencies between the software and the associated hardware and firmware give rise to technical dependencies, arising as a result of dependencies in software code. For example, some security vulnerabilities are present in multiple functions in the source code that have interdependencies among them. Consequently, security patches developed to fix them also include these function-level dependencies that must be carefully handled during patching. \\

In terms of \textbf{software-related patch dependencies,} several factors such as operating system (OS) dependencies, software application dependencies, and pre-requisites for patch installation fostered the conditions for coordinating the dependencies. Patch management in large software systems involves managing multiple software components (or services) with different OS versions. The existence of the interdependencies between OS versions and other software applications built on top of a particular OS like web browsers may create additional tasks for practitioners as all of them need to be synchronized. 

\hspace{1 mm} \textit{``We have about 15-16 versions of Windows 10. So, before patching we need to see which version is running on which server? What is the build number? Are we running the latest? It's a lot!"} - AT4, M14-dis

\hspace{1 mm} Additionally, some security patches contained interdependencies with legacy OS. It presented a much more arduous task to the studied practitioners since some critical emergency medical services were running on Alpha's legacy systems that were not supported by large vendors, e.g., Microsoft. In such cases, the participants often felt forced to delay patch installation as the available solutions like decommissioning or upgrading legacy systems presented high risks of operation interruptions. Consequently, this practice of delayed patching of security vulnerabilities in systems would significantly increase the risk of exposure to attacks. 

Similarly, interdependencies between software applications, platforms, and tools presented another major category of software-related dependencies. 
This is because the build dependencies between the software application and patch sources require the versions to be in sync before attempting patch installation \cite{duan2019automating}. As the size of an organisation grows, managing these dependencies appeared difficult with a large number of diverse applications installed. For example, Alpha's software applications ranged from general applications, e.g., Java, .NET to specific medical applications, e.g., Electronic Medical Record Software. As such, Beta teams spent most of their time detecting the existing interdependencies such as version incompatibilities between various software applications. 
    
    
\hspace{1 mm}\textit{``There’s an old HP tools version and a new version, and the vulnerabilities are coming up on the scan as with the new version. But the issue is because the old version is still there which we should have got rid of earlier."} - BT1, M14

\hspace{1 mm} The circular dependency represented a more complex semantic dependency in security patch management. An example scenario was when software B is dependent on software A, and software A uses software B to function (B $\leftrightarrow$ A). In such cases applying a security patch to A led to service unavailability of B as a result of rebooting A to make the security patch take effect. In particular, effectively coordinating circular dependencies was crucial in the healthcare domain as A and B could represent critical medical services like emergency life support or surgery equipment. 

\hspace{1 mm}\textit{``And there could be like A needs B to run, and vice versa but when we accidentally took B offline that day, A didn’t work. That was when we all got goosebumps."} - AT4, M10-dis
    
On the other hand, some security patches required pre-requisites to be established before installation for the patch to take effect. In most cases, the pre-requisites comprised registry changes and preparation package installation. We identify that it is resulting from the patches that do not contain source code modifications, as explained by Li and Paxson \cite{li2017large}. To investigate the pre-requisites of the security patches retrieved each month, Beta allocated a specific timeframe before patch testing and discussed with Alpha during the patch meetings how and when they would handle the identified pre-requisites. Coordinating the pre-requisites was often a manual task as it involved decisions about suitable configurations based on the organisation's needs and the other associated software dependencies. Failing to configure the pre-requisites led to errors that would halt a patch installation. However, we found that the teams became more receptive to detecting pre-requisites-related dependencies with the continuous early investigation approach employed.
    
\hspace{1 mm}\textit{``The patches listed here needed a preparation package installed before the patching window and then the reboot would have applied the patch. We’ll do that just before the current patching window and then patching should proceed as normal without errors."} - BT1, M11 \\
    

Besides the most common software-related dependencies, some security patches also contained dependencies with the associated hardware and firmware giving rise to \textbf{Hardware and Firmware-related dependencies}. For example, in one instance, practitioners were unable to patch the security vulnerabilities found in virtual machine (VM) software as the VM-related firmware was not up to date. So, they had to regularly keep track of the existing dependencies and update the supporting hardware and firmware accordingly before attempting patch installation.


\hspace{1 mm}\textit{``Some patches need a certain type of hardware to be at a certain level. There was a 2008 security patch which we couldn’t install until we updated the firmware or the utilities."} - AT4, M10-dis

\subsubsection{\textbf{Social dependencies}}

\begin{figure*}[h]
  \centering
  \includegraphics[width=0.75\linewidth]{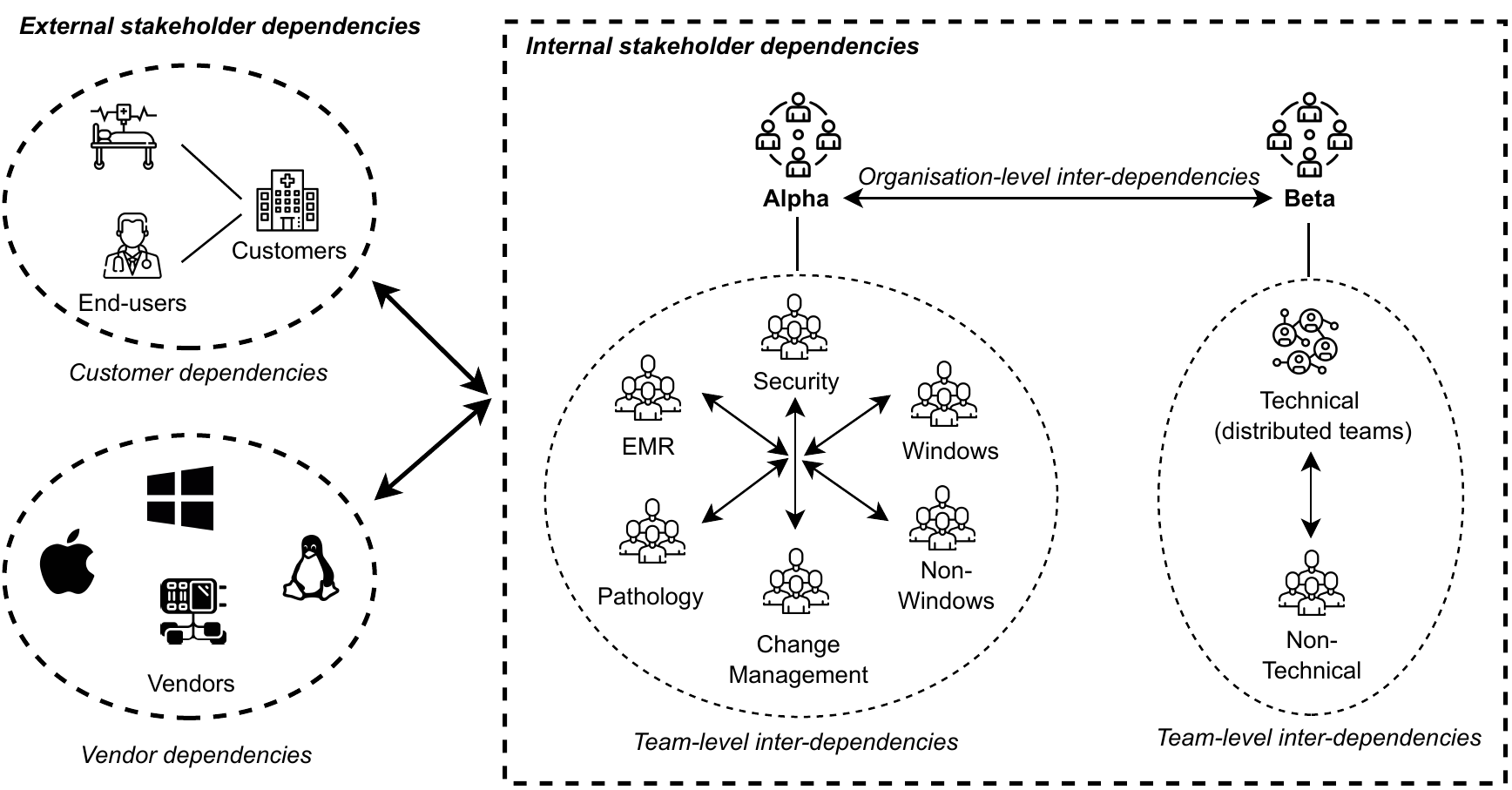}
  \caption{The social dependencies present in the studied context.}
  \Description{The stakeholder dependencies present in the studied context of Alpha and Beta.}
  \label{fig:socialdependencies}
\end{figure*}

Social dependencies that stem from interdependencies between stakeholders is another major category of dependencies integral to security patch management. Security patching in large organisations is challenging due to the increased complexity stemming from a high number of stakeholders. Therefore, effectively coordinating the dependencies between multiple stakeholders is important for successful patch management. Our analysis of the gathered data led to the emergence of two sub-categories of social dependencies: \emph{Internal stakeholder dependencies} and \emph{External stakeholder dependencies} as illustrated in Figure \ref{fig:socialdependencies}. \\

\textbf{Internal stakeholder dependencies -} The two organization's stakeholders worked together to achieve a common goal of securing the state's healthcare system by timely installation of security patches. Hence, the internal stakeholder dependencies, in this context, relate to the dependencies stemming from the interactions between stakeholders across Alpha and Beta. We identified two layers of dependencies namely \emph{Team-level dependencies} and \emph{Organisation-level dependencies}. 

The current context displays a \emph{multiteam system} (MTS) structure \cite{mathieu2001multi} as multiple interdependent teams within each organization collaborated towards a collective goal. The interdependent team interactions gave rise to \emph{team-level dependencies.}
Since each organisational team had assigned responsibilities, coordination between and within teams remained pivotal to achieving the goals. In most cases, the inter-team tasks contained dependencies that required management of the cross-team interconnections. For example, Alpha teams T1, T2, and T3, often depended on T5 to approve security patch schedules before assigning them to Beta to be executed. Lack of awareness of the roles and responsibilities complicated the coordination of team dependencies causing delays of several weeks in patching known vulnerabilities. 

\hspace{1 mm}\textit{``We are still waiting for an email from [P1] approving the manual patching process. (BT1)\\
Well, I’m not sure whether it should come from [P1] or some other guy. I will confirm it with [P2] and get to back you."} - AT2, M8
    
Similar to the inter-team dependencies, \emph{organisation-level dependencies} also created several challenges for security patch management. The large scale and heterogeneous nature of Alpha created additional complexities to coordinating organisation-level dependencies that often resulted in delays in patch testing and deployment. For example, every security patch required approval from Alpha's Change Management (T5) before they could be installed at the customer's sites (i.e., hospitals). Given the critical nature of the task, it was important to effectively coordinate it and leave sufficient time for the customers' agreements. It was needed to reduce the time of service interruptions and manage technical dependencies.

\hspace{1 mm}\textit{``This morning [Alpha's] change manager said that the change hasn’t been approved yet. So, we had to suddenly change plans just minutes before the [scheduled] patch window."} - BT1, M10 \\

\textbf{External stakeholder dependencies -} The involvement of external stakeholders such as customers, end-users, and vendors is integral to security patch management. Effectively coordinating the collaborative relationships and the dependencies with external stakeholders is vital to a sound patch management process. In this context, the external stakeholders consisted of customers (e.g., hospitals) and vendors (e.g., Microsoft). The end-users dependencies consisted of hospital patients and staff, hence, these were not directly linked to Alpha and Beta. However, ineffective coordination of customer dependencies negatively impacted end-users. For example, uninformed operation interruptions to medical equipment resulted in inconveniences to patients and medical staff. In contrast to internal stakeholder dependencies, managing external stakeholder dependencies presented a much more difficult challenge to practitioners. The main reason was the lack of a shared understanding of the importance of security patching and the visibility of the existing process interdependencies.

\emph{Vendor-related dependencies} refer to the dependencies that are created due to the need of installing security patches received from vendors. Management of vendor dependencies became difficult with the presence of shared vulnerabilities and associated technical dependencies in software applications. This demanded coordination of patch releases from the vendors of different software applications. Additionally, some of Alpha's third-party applications were patched by external vendors, for example, the medical application providers, as per the agreement at the point of purchase. Thus, it required synchronising each vendor's patching cycles to avoid the unavailability of the interdependent systems.  

\hspace{1 mm}\textit{``Regarding the recent concern from Security (T4) on [S1 server] patching is one month behind, can we confirm this with [third-party vendor's] requirements? Because this vendor’s patching cycle is always one month behind, every month they release the patches for the last month."} - BT1, M17

\hspace{1 mm} Furthermore, missing, faulty, or exempted security patches and unknown errors during patch installation increased the need for coordinating the vendor dependencies during patch testing, deployment, and post-deployment patch verification. Patch exemption was when selected security patches were excluded from installation due to legitimate reasons approved by Alpha's Security team. We observed several scenarios of patching delays due to a lack of coordination of software dependencies with exempted security patches that were managed by external vendors.
    
Managing \emph{customer-related dependencies} with hospitals presented a challenge to Alpha and Beta teams, particularly when negotiating the patch schedules. Reaching a consensus on a patch installation time was essential to minimize any potential impact of service disruptions from reboots. However, an interesting observation was that in a majority of cases, practitioners spent most of the time trying to communicate to customers on the need to patch systems, as opposed to agreeing on the patch installation time. It was due to the lack of understanding of the need to apply security patches and the inability to accept the high risks of service downtime.
    
\hspace{1 mm}\textit{``A lot of customers don’t always understand the worth of security patching, they just want to use the server, and keep asking; ``why do you want to reboot it every then and there, or why you got to update it? It’s working and leave it alone!"} - AT4, M4-dis
    


\subsection{Constraints}

This section presents the constraints that hindered the coordination. When the constraints affected the socio-technical dependencies, they caused coordination breakdown. Hence, it is important to devise suitable approaches and tools for identifying and managing the potential impact of constraints. 

\subsubsection{Legacy software-related dependencies}

Legacy systems pose a security threat to organisational ICT infrastructures, particularly in mission-critical domains like healthcare. This is because most of the critical services that run on legacy systems remain unsupported by vendors leaving them vulnerable to security attacks. Alpha had several legacy systems like 2008 servers that were no longer receiving security patches from Microsoft. Furthermore, the dependencies with these legacy systems produced wider implications for security patching. It resulted in the practitioners being unable to patch until the dependent legacy systems were upgraded to the current version or offered extended support from vendors. However, upgrading legacy systems required a critical evaluation of the impact on other important services. In some cases, Alpha had acquired extended support for critical legacy systems through negotiations with the vendors. However, it presented practitioners with additional challenges of having to perform manual configurations to install security patches as the current configuration settings did not work, and some installed patches getting rolled back. 
    
\hspace{1 mm}\textit{``There are twenty six 2008 servers still waiting to be patched this month. But there are some servers that we need to have a look at why the patches aren’t applying. Even though we have installed all the required preparation packages, they keep rolling back."} - BT1, M12

\subsubsection{Lack of automation support} 

Lack of suitable tool support presented a major constraint for the coordination of the dependencies. One of the key constraints was the inability to oversee technical dependencies across all systems inventory. Inability to identify the existing software dependencies from the current tools resulted in practitioners spending hours trying to find the current software versions in the event of errors during a patch installation. Additionally, the lack of automation support to investigate the patch pre-requisites as mentioned in Section 4.1.1, caused delays and sometimes errors in installing patches due to missing out on some registry changes.
    

\hspace{1 mm} On the other hand, the limitations on the features of the available support tools presented constraints in detecting specific dependencies such as legacy dependencies and their contextual categorization. This could be because most of the tools available to the studied teams focused on function-level patching assuming that vulnerable code resides within only one function \cite{duan2019automating, li2017large}. For example, Alpha's decisions were largely based on the vulnerability scan reports. However, the existing scanning tool was unable to filter the unpatched vulnerabilities resulting from legacy dependencies and exempted patches, which constrained accurate decision-making.

\hspace{1 mm}\textit{``These vulnerability numbers will go down by as much as half since [scanning tool] captures 2008 servers' vulnerabilities as well. I don’t know whether we can do exceptions through the tool, like flag things that are legacy, to make the numbers reflect what we see."} - BT1, M12

\subsubsection{Increased patch load} 

As the organisation size grew, the number and diversity of systems also expanded that resulted in increased complexity in the patching. Hence, the practitioners often faced difficulties in keeping up with the patch release rate. Accumulated patch load due to the previous patch exclusions added to the challenges as they had to patch previously excluded patches in the following month. Correspondingly, an increased patch load led to more socio-technical dependencies creating additional constraints on coordination. Overall, it led to an increased risk of exposure to attacks as installing security patches was often delayed.

\hspace{1 mm}\textit{``There’s just too much to check! We’re dealing with 1500 servers, we don’t have time to look at each patch for every server like, ``Yeah, this one is right, this isn't. Which servers have interdependencies that can't be patched at the same time? Which server has which version of this software, that software..."} - AT4, M14-dis

\subsection{Breakdowns}

In this subsection, we report the scenarios that exemplify the breakdowns resulting in the security patch management process from ineffective coordination of the socio-technical dependencies and the related constraints.

\subsubsection{Sudden escalations to patch schedules}

Security patch installation within the allocated patch window is critical in a mission-critical domain like healthcare to avoid unexpected service disruptions. Alpha provided specific patch windows to Beta, usually 4-hours, to install the security patches on production servers (i.e., operating in hospitals) after agreeing with the customers. To adhere to the specified patch window while installing patches, Beta teams were required to plan well ahead to establish each patch's pre-requisites, identifying interdependencies, testing the patches, and obtaining management approval. As such, miscoordination of these conditions would usually lead to unexpected escalations to the planned patch schedules. Given the critical nature of the healthcare operations, security patch installations going beyond the scheduled patch window resulted in devastating consequences such as life-threatening risk to critical patients from additional service unavailability even for a couple of minutes.

\hspace{1 mm}\textit{``We realized there's a need to do a sudden change of configuration for [ISP] servers at the time of patching, so our team had to escalate immediately to switch to manual patching because some servers needed Windows approved patches to be rolled out."} - BT1, M12

\subsubsection{Delays in the organisation approvals}

All security patches needed to be approved by Alpha's change manager a month before a patch installation to allow Beta teams to prepare in advance for installing a patch. However, we observed some delays in approvals from the Change Management team resulting in abrupt changes to patch installation plans such as shorter patch windows. Such unforeseen changes invoked changes to the dynamics of socio-technical dependencies resulting in breakdowns in the process. For example, having to install the emergency security patches in shorter patch windows warranted re-testing of patches to confirm that patch installation can fit into the shorter patch window, and obtaining reapproval from customers to avoid unexpected service interruptions.   

\hspace{1 mm}\textit{``We were ready for the patch deployment. But this morning [Alpha's] change manager said that the change hasn’t been approved yet. So, if it isn’t approved prior to scheduled start time, we will have to reschedule it."} - BT1, M10

\subsubsection{Lack of dependency awareness from localized work distribution}

Alpha and Beta teams had localized work distribution settings within their team structures. Alpha teams were located in the same office and Beta teams were distributed across different offices in the same city. This structure led to the creation of a lack of dependency awareness of the task assignments and progression between teams. We observed this during status reporting in the meetings as some members were unaware of the tasks progressing in the teams. Their lack of understanding resulted in added challenges of coordinating the inter-team dependencies that inhibited measuring the progression of security patch management tasks. 

\hspace{1 mm} \textit{``I believe [P1 from AT2] is working on this issue at the moment. Do you know when he’s likely to get that done? - (BT1) \\Not sure, but I would be hoping next week. I would contact him and let you know."} - AT2, M13

\subsection{Mechanisms}

This subsection presents a collection of strategies that have emerged from analyzing the data; the studied teams practiced these mechanisms to manage the dependencies while mediating the constraints.
 
\subsubsection{Early investigation of interdependencies}

Alpha and Beta teams used patch meetings to discuss the findings from the investigations of the technical dependencies. Since the participants with diverse technical backgrounds and expertise attended the meetings, this configuration enabled knowledge sharing to collaboratively identify dependencies upfront. Failing to identify the dependencies resulted in scenarios of installed patches not working as intended, patches refusing to get installed resulting in rollbacks, and patches going beyond the allocated patch window during installation. Early identification of dependencies helped practitioners to coordinate the task dependencies among teams, and make timely decisions to address the usual problems such as raising support cases to vendors seeking expert advice and finding workaround solutions. 

\hspace{1 mm}\textit{``From troubleshooting why the last two weeks' patches hadn’t worked properly, we can realize that each patch needed to be rebooted at the beginning. Since we failed to do so, the current reboot may have applied the last month's patches."} - AT4, M14

\subsubsection{Collaborative decision-making}

Accurate and timely decision-making is pivotal throughout the patching process. The studied practitioners used patch meetings as a platform to collaboratively decide about vulnerability risk assessment and prioritization, and approval of the patch decisions and schedules. Collaborative decision-making helped the teams to maintain dependency awareness about the decisions and to plan the associated tasks with minimum impact of dependencies. For example, the decisions about patch exemptions involved a collective assessment of the requests. Thus, the awareness of patch exemptions helped the teams to plan their other patch schedules with limited dependencies to the exempted patches, and keep track of the exempted patches in the month and organise the to-do patch list in the following month including them.

\hspace{1 mm} \textit{``The last item on the agenda is about the [s1] servers that are marked as excluded from patching. We need to decide if they are being exempt from our patching list this month or not because we haven’t got the official confirmation whether we're doing the patching or if the [third-party] vendor is doing it again?"} - BT1, M13 


Other examples of collaborative decision-making involved selecting the optimum patch configurations based on organisational needs, and managing legacy software dependencies. In most cases, the decisions for legacy software dependencies revolved around the need for decommissioning or rebuilding legacy systems, when and how to do it, and how to patch them following the rebuilds. Decisions about the patch schedules for the approved patches helped the teams to coordinate the planning upfront and identify the need for \emph{out-of-band (OOB) patching}. OOB patching refers to the need to allocate an additional patch window when some security patches require more installation time than the allocated patch window due to compound dependencies involved.  


\subsubsection{Continuous measuring of progression}

Alpha teams measured the continual progress of vulnerability remediation through Beta's status reports in meetings and regular vulnerability scan reports. When the scans indicated an increase in the number of vulnerabilities present in the systems, the matters were discussed extensively to remedy the situation. The continuous measurement of the progression enabled the identification of the outliers such as missing patches resulting in the investigation of the causes and coordinating the associated stakeholder dependencies with the third-party vendors.

\hspace{1 mm} \textit{``What is the status of internal [s1] server security vulnerabilities? - (AT4) \\
We’re getting our regular scans to measure that. That one is progressing quickly. We will share the report next week"} - BT1, M9

\subsubsection{Frequent communication}

Frequent communication appeared to be essential for effective coordination of dependencies. It helped to erase boundaries between roles, teams, and organisations, and increase cohesion and trust between stakeholders. Teams used various communication mediums such as bi-weekly patch meetings, email, and Skype. Additionally, the studied practitioners held separate meetings to discuss critical and urgent matters that emerged in between patch meetings or when patch meeting discussions were dragged beyond the allocated time. Patch meetings were the most preferred communication medium as the teams felt more comfortable with direct communication.
Communication during patch meetings facilitated collaboration, knowledge sharing, and information exchange about technical and socio-technical matters affecting the patching process, for example, upcoming patch schedules, changes to patching plans such as out-of-band patch schedules, and vendors’ patch release information. Regular patch meetings benefited the teams in numerous ways such as allowing visibility into task progression and assignments, staying proactive to potential issues about critical security vulnerabilities, and effectively coordinating security patch management activities. 

\hspace{1 mm}\textit{``[Security Advisor shares the vulnerability remediation progress report on screen] We were averaging 75 high-risk vulnerabilities per server back in 2016 when I joined. As you can see now, we’re down to 5 per server. Given the mix of environments we are dealing with, this is amazing. You can see that the frequent patch meetings making a big difference!"} - AT4, M10

\subsubsection{Load balancing}

An interesting strategy employed by the studied teams to coordinate patch schedules was load balancing. It was used to balance the patch load in servers at any given patch installation time. Balancing the server load helped reduce service interruptions. Patching dozens of servers at the same time significantly increased the risk of system failure as all the servers go offline at the same time during reboots. Load balancing, on the other hand, helped to run the critical medical services concurrently on another server while the desired server(s) is being rebooted. However, the presence of technical dependencies created difficulties in load balancing. In particular, for instances with one-to-one dependencies such as (A $\rightarrow$ B), the practitioners had to rigorously analyse the interdependencies before planning the load on servers to avoid unexpected system downtime. 

\hspace{1 mm}\textit{``Before we started with the load balancing, we patched 50 servers one night, and just two the next night. So, I suggested we plan to load balance. But there's a lot to manage, especially when we have systems like, system A is redundant to B, and oops! we accidentally took both of them down at the same time to patch."} - AT4, M10-dis

\subsubsection{Centralized vulnerability risk assessment}
    
Regularly performing vulnerability risk assessment and prioritization was necessary as it could potentially differ from that of the vendor's assessment based on the organisation's environment. It aided practitioners to plan well in advance to promptly respond to critical security vulnerabilities. To regularly monitor security vulnerabilities, the teams had devised a centralized role in Alpha's Security team (Security Advisor) responsible for scanning and categorizing the vulnerabilities based on teams' ownership. Having a centralized structure helped maintain consistency in vulnerability risk assessments across teams as well as reduce delays in vulnerability assessment and prioritization decisions. Additionally, frequent comparisons with the previous scans assisted with evaluating the vulnerability remediation performance.
    
\hspace{1 mm}\textit{``[Security Advisor] gets the global rating of a vulnerability risk and re-assess it to see if it's critical to us and how it can be exploited. For medium to low risks, we patch in the next cycle, but if it’s critical or we’re under attack, we’ll patch within 48 hours."} - AT4, M10-dis

\section{Discussion}
\label{section:discussion}

In this section, we discuss a comparison of our theory with the prior related work, elaborate on the broader implications of our theory for practitioners and researchers, and reflect upon the potential threats to the validity of this study. \\

\textbf{Comparing to Related Work:} Following Glaser’s advice \cite{glaser1967discovery, stol2016grounded}, we compare our theory with the existing literature. The prior related work on this topic \cite{nappa2015attack, li2019keepers, huang2012patch, potter2005reducing, dissanayake2020software} primarily focuses on the coordination and collaboration challenges but does not provide in-depth details of the causes or the potential strategies to address the coordination challenges leading to delays in applying security patches. A few studies \cite{nicastro2003security, prochazka2011race, hanauer2018process} investigate the social dependencies concerning the involvement of multiple internal and external stakeholders. Another set of studies \cite{nappa2015attack, cavusoglu2008security, dey2015optimal, cavusoglu2006economics} exclusively focus on vendor dependencies that might arise from shared vulnerabilities in software code. In particular, their main focus remains on optimizing patch management by obtaining an equilibrium of an organisation's patch cycle with a vendor's patch-release cycle to minimise cost. An important observation is the absence of \emph{theories} that focus on the socio-technical aspects concerning security patch management in contrast to quantitative \emph{models} and \emph{frameworks} \cite{ralph2018toward}. This is an important point as the theories provide \emph{``basic concepts and underlying mechanisms, which constitute an important counterpart to the knowledge of passing trends”} \cite{hannay2007systematic}. In contrast, our theory derived from the gathered data differs from these existing works in several ways as it:\\
 • explains the role of coordination in security patch management as incorporating a multi-dimensional nature across four inter-related dimensions in contrast to focusing on one type such as vendor dependencies reported in \cite{nappa2015attack, cavusoglu2008security, dey2015optimal, cavusoglu2006economics}, \\
 • explains the socio-technical dependencies that create the need for coordination going beyond reporting just the challenges with lack of coordination \cite{nicastro2003security, prochazka2011race, hanauer2018process}, \\
 • explains the constraints that hinder coordination and shows scenarios of breakdowns resulting from ineffective coordination, \\
 • suggests strategies that can be used for effective coordination, \\
 • offers a comprehensive overview of the impact of coordination in security patching in the mission-critical domain like healthcare, \\
 • presents a theoretical model for future research, and, \\
 • provides guidance to practitioners to overcome patching delays, and increase confidence in their decisions. \\

\textbf{Implications for Practitioners:} The reported theory can be used to gain an in-depth understanding of the significance and the impact of the role of coordination in security patch management. Practitioners can use this understanding to realize their roles and responsibilities in ensuring coordination effectiveness across different dimensions of the role of coordination. Moreover, practitioners can use the theory as a guide to identify the related dependencies and how they might affect their security patching process. Further, we have observed that adopting these coordination mechanisms has resulted in a reduction of the unpatched security vulnerabilities in Alpha systems. Hence, our findings may also be useful in exploring the suggested mechanisms in their organisational setting. Additionally, practitioners can benefit from the early detection of constraints and breakdowns to avoid failures.  \\

\textbf{Implications for Researchers:} 
Given a Grounded Theory study is considered to produce a ``mid-ranged” theory based on the contexts studied \cite{glaser1992basics}, other researchers can carry out an extension through future research including a more detailed analysis of the present dimensions, new dimensions discovered, or different contexts. Context-specific research investigating how the role of coordination is impacted by contextual factors can result in useful models of coordination in patch management \cite{cheng2016context, rodriguez2020theory}. For example, organisation-level dependencies may not directly apply to small organisations where security patch management is usually handled by one team. Future studies can also investigate the effectiveness of the coordination mechanisms and the context in which they should be employed. The impact of organisational policies is often cited as one of the dominant socio-technical challenges in security patch management \cite{li2019keepers, tiefenau2020security, nicastro2003security, prochazka2011race, yang2011sla, dissanayake2020software}. Similarly, future studies can explore how organisational culture affects the role of coordination. Another possibility is to employ the findings in large-scale surveys to evaluate the theory and identify variations in different organisation settings such as in DevOps processes. 

While this study is based on data collected from software security patch management, the findings can be directly beneficial to the software development research. This is because patch application is inherently dependent on patch development. An important point to note from our theory is the need to consider the \emph{socio-technical} aspect intrinsic to patch management when developing patches (e.g., future work similar to Li et al. \cite{li2019keepers}). We show that early identification of the dependencies is the key to avoid patching delays and failures, but lack of automation support presents a key constraint as previously mentioned for timely identification of the dependencies. Therefore, there is a need for research about how Artificial Intelligence (AI) support can be employed in dependency detection in patch development and management. The findings also highlight the important need for further R\&D for advanced patch management tools. For example, scanning tools can be enhanced to customize the software dependencies such as excluding exempted patches to provide real-time feedback that assists practitioners with accurate decision-making. Furthermore, our theory provides an in-depth understanding of how the role of coordination impacts the mission-critical domain, particularly, healthcare. An understanding of the causes of unexpected service interruptions can help researchers to devise strategies to avoid such downtime. While the research into dynamic software updating \cite{hicks2005dynamic} (DSU) attempting to address this issue is progressing, our results can be useful information for future research that investigates the effectiveness of the developed strategy in mission-critical contexts. \\

\textbf{Threats to Validity:} A Grounded Theory study does not affirm generalization as the theory formulation is pertinent to the studied context \cite{glaser1967discovery, glaser1978theoretical}. The context of this study is limited to the cases studied in security patch management in the domain of healthcare. Nevertheless, we believe that our theory can be recreated in other contexts and modified. 

In terms of data representativeness, our data collection is limited to the observations of patch meetings, post-meeting discussions, and analysis of meeting minutes and patch mailing thread as described in Section 3.1. We acknowledge that more data sources such as interviews and surveys can be incorporated in future studies to increase the scope of the analysis and verifiability of the theory \cite{silva2020theory}. 

While employing Grounded Theory procedures permits the data analysis to be grounded in collected data, there is a threat of subjectivity of the data analysis referred to as the ``uncodifiable step" \cite{langley1999strategies, baltes2018towards}. To alleviate this threat, we regularly held internal discussions on the emergent findings throughout the study as described in Section 3.2. In addition, the findings were further cross-checked with a senior member of Alpha’s security team to ensure we have accurately interpreted the theory from the observed practices. 

The verifiability of a grounded theory can be deduced from the robustness of the research method, and evidence of theory formulation from its application \cite{glaser1992basics}. To confirm verifiability, we have described our application of the Glaserian version of GT in detail (Section \ref{section:method}) and included quotations from the underlying data in the findings (Section \ref{section:findings}). These details provide evidence of how our theory meets the GT evaluation criteria: the generated categories \emph{fit} the underlying data (see Figure \ref{fig:researchmethodcoding}); the theory can \emph{work} as it explains the main concerns of the participants in patch meetings; 
the theory has \emph{relevance} to the domain of software security patch management; and the theory is open to \emph{modification} based on future studies in different contexts \cite{glaser1978theoretical, stol2016grounded}.

\section{Conclusion}
\label{section:conclusion}
We present the Theory of the Role of Coordination in Software Security Patch Management. The developed theory explains the effects of coordination in the patch management process across four inter-related dimensions namely \emph{causes, breakdowns, constraints,} and \emph{mechanisms}. Our theory is based on a longitudinal Grounded Theory study of 51 patch meeting observations involving 21 industry practitioners in two organisations in the healthcare domain over a duration of 9 months. We provide the grounded evidence that the role of coordination represents a core concern, contrasting with a perception among the SE community that automation and tooling alone can be sufficient to achieve success in patching and highlight the need to have a delicate balance between the socio-technical concerns such as coordination and automation to reduce delays, which is often unrecognised in the existing literature.

Overall, besides providing a holistic understanding of the role of coordination in security patch management that is based on empirical evidence and grounded in practice, our study is the first attempt to investigate in-depth the socio-technical aspects of security patch management in the mission-critical healthcare domain. The theory provides important insights for practitioners to avoid patching delays and failures and enhance confidence in their decisions, and for researchers to shape their work on patch development to address the practical concerns in patch application. The findings can also be used for developing the next generation of AI-enabled tools for supporting the patch management process.


\newpage
\begin{acks}
We would like to thank the practitioners and organisations that collaborated with us for their valuable support. This study was conducted under the University of Adelaide Human Research Ethics Committee Application ID H-2020-035.
\end{acks}

\appendix
\section{Observation protocol}
\label{appendix:observationprotocol}

\begin{table}[ht]
  \caption{Observation protocol}
  \label{tab:appendix}
  \centering 
  \small
  \begin{tabular} {p{.1\textwidth} p{.33\textwidth}}
    \toprule
    Topic & Question\\
    \midrule
    Participants & What are the roles and other details of the participants? \\
      & \hspace{0.3cm}• Number of attendees \\ 
      & \hspace{0.3cm}• Role (manager, system administrator, etc.) \\ 
      & \hspace{0.3cm}• Affiliation (Alpha or Beta) \\ 
      & \hspace{0.3cm}• Team (Security, Windows, Server, etc.) \\ \\
      & Is someone acting as a facilitator? \\
      & \hspace{0.3cm}• Who? \\ 
      & \hspace{0.3cm}• How is he/she facilitating the meeting? \\ \\
    Communication & What is the communication channel? \\ \\
      & How does the communication happen? \\
      & \hspace{0.3cm}• Directed / indirect questions? \\ 
      & \hspace{0.3cm}• Active participation in communication? \\ 
      & \hspace{0.3cm}• Any roles that are most active in communication? \\ \\
    Activities & What are the various discussions and activities? \\
      & \hspace{0.3cm}• Topics discussed \\ 
      & \hspace{0.3cm}• Challenges discussed \\ 
      & \hspace{0.3cm}• Activities (demonstrations etc.) \\ \\
    Objects & What resources/media are used? \\
      & \hspace{0.3cm}• Presentation slides, excel sheets, tools, etc. \\ \\
    Collaboration & How do the participants interact and corporate with each other? \\ \\
    Events & Are there any particular events or anything unanticipated? \\ \\
    Time  & When does the meeting start? \\
      & What is the sequence of events? \\ 
      & When does the meeting end? \\ \\
    Goals  & What are the participants trying to accomplish? \\ \\
    Feelings & How is the atmosphere? \\ \\
    Closing & How is the meeting ended? \\
      & Is there a post-meeting planned? \\ 
      & Is there anything discussed about the next meeting? \\
    \bottomrule
  \end{tabular}
\end{table}

\clearpage 
\bibliographystyle{ACM-Reference-Format}
\bibliography{bibliography}

\end{document}